\documentclass[aps,pre,superscriptaddress,preprint,amsfonts,amssymb,amsmath,bibnotes,footinbib,showkeys,showpacs,titlepage]{revtex4}

\usepackage{graphicx,color}
\usepackage{ifpdf}
\usepackage[latin1]{inputenc}
\usepackage{float}

\begin{document}

\title{Analysis of an information-theoretic model for communication}

\author{Ronald Dickman}
\email[e-mail: ]{dickman@fisica.ufmg.br}
\affiliation{Department of Physics of the Federal University of Minas Gerais,\\
Belo Horizonte, Minas Gerais, Brazil}
\affiliation{National Institute of Science and Technology for Complex Systems,\\
Caixa Postal 702, 30161-970 \\ Belo Horizonte, Minas Gerais,
Brazil}
\author{Nicholas R. Moloney}
\affiliation{Department of Physics of the Federal University of Minas Gerais,\\
Belo Horizonte, Minas Gerais, Brazil}
\affiliation{National Institute of Science and Technology for Complex Systems,\\
Caixa Postal 702, 30161-970 \\ Belo Horizonte, Minas Gerais,
Brazil}
\author{Eduardo G. Altmann}
\affiliation{Max Planck Institute for the Physics of Complex Systems,
01187 Dresden, Germany}

\date{\today}

\begin{abstract}
We study the cost-minimization problem posed by Ferrer i Cancho and Sol\'e in their
model of communication that aimed at explaining the origin of Zipf's law [PNAS 100, 788 (2003)].
Direct analysis shows that the minimum cost is
$\min \{\lambda, 1-\lambda\}$, where $\lambda$ determines the relative weights
of speaker's and hearer's costs in the total, as shown in several previous works
using different approaches.
The nature and multiplicity of the minimizing
solution changes discontinuously at $\lambda=1/2$, being qualitatively
different for $\lambda < 1/2$, $\lambda > 1/2$, and $\lambda=1/2$.
Zipf's law is found only in a vanishing fraction of the
minimum-cost solutions at $\lambda = 1/2$ and therefore is not explained by this model.
Imposing the further condition of {\it equal costs} yields distributions
substantially closer to Zipf's law, but significant differences persist.
We also investigate the solutions reached by the previously used
minimization algorithm and find that they correctly recover global minimum
states at the transition.
\end{abstract}


\maketitle

\section{Introduction}

\vspace{1.5em}

Among the numerous empirically reported power-law distributions, one of the oldest and with best statistical
support is Zipf's law, which
states that the frequency $P(k)$ of the $k$-th most frequent word decays as
$P(k) \approx C/k^\mu$ with $\mu \simeq
1$~\cite{zipf}. While there are various stochastic models of text
generation that reproduce this and other statistical features of
corpora~\cite{Newman,Baayen,zanette}, a definitive answer to the more fundamental
question of why natural language shows Zipf's law is still lacking.
Zipf argued that it is a consequence of the tendency of speakers and hearers to communicate
with least effort~\cite{zipf}.
Ferrer i Cancho and Sol\'e  recently proposed a quantitative model that builds
on these ideas and suggests how natural language could
have evolved to a state satisfying Zipf's law~\cite{FCS}.
The importance of this work, which we revisit here, is that it introduced
a framework of language games
to explain Zipf's law which influenced many subsequent works~\cite{FC,FCDZ,trosso,PAOP}
and contributed to the current interest in modeling
different aspects of language dynamics~\cite{Baronchelli}.

In the framework introduced in Ref.~\cite{FCS}, which fits into a more general modeling scheme
of language evolution~\cite{Komarova,Nowak}, Ferrer i Cancho and Sol\'e
considered a scenario of ``objects" reported to a listener by a speaker
using a certain lexicon of symbols. The speaker's cost of communication,
$\alpha$, is related to the average information per symbol, and the listener's
cost, $\beta$, to the mean uncertainty associated with the of symbols.
(Uncertainty arises when a symbol denotes more than one object.)
It was proposed that as the language evolves the
cost function $\Omega(\lambda) = \lambda \beta + (1-\lambda) \alpha$ is minimized,
with the parameter $\lambda \in [0,1]$.  Studying the minimization problem numerically, the authors
assert that the model exhibits a phase transition at a certain value of $\lambda$, at
which Zipf's law is satisfied.
Continuous phase transitions are related to power-law distributions (associated, for example, with
long-range correlations and self-organized criticality) and their possible
connection to Zipf's law is another appealing idea of Ref.~\cite{FCS} that motivates our work.

In \cite{FCDZ,trosso,PAOP} it was shown that the
minimum cost in the language game proposed in Ref.~\cite{FCS}
is simply $\min \{\lambda, 1-\lambda\}$.
In this paper we demonstrate this result in a simple manner starting from
the inequality $\alpha + \beta \geq 1$, and investigate
various aspects of the model.
At $\lambda = 1/2$ the nature of the minimum-cost
state changes discontinuously and multiple minimum-cost states with
varied properties coexist. The states satisfying Zipf's law are found to be extremely
rare, comprising a vanishing fraction of all
minimum-cost states in the relevant limit of large number of symbols and objects.
We also apply a {\it numerical} minimization approach to
see if the observation of Zipf's law can arise from a
failure to attain the minimum-cost state (as speculated in Ref.~\cite{FCDZ}).
Using a stochastic algorithm along the lines proposed in
\cite{FCS}, we verify that minimum-cost states are indeed
attained at the transition, but that are they are non-Zipfian, that is,
the associated rank-frequency distribution does not exhibit a broad region with power-law decay.

The remainder of this paper is organized as follows. In Sec.~\ref{sec.model} we define the model
in detail, in Sec.~\ref{sec.states} we determine the states minimizing the cost function
$\Omega(\lambda)$, examine their properties, and investigate the consequences of the
equal-cost condition discussed in \cite{PAOP}.  In  Sec.~\ref{sec.simulation} we report the
results of the simulations, and in Sec.~\ref{sec.conclusions} we summarize our conclusions.

\section{Model}\label{sec.model}

In this section we define the model proposed by Ferrer i Cancho and Sol\'e using a notation
and terminology that differs somewhat
from that of \cite{FCS}, but which we believe facilitates the analysis.
Consider the interaction between a ``speaker" and a ``listener"
in a language consisting of $n \geq 1$ symbols, $s_1,...,s_n$,
used to describe a world of $m \geq 1$ objects, $r_1,...,r_m$.
The relation between symbols and objects is defined via a {\it lexical matrix} {\sf A}~\cite{Komarova,Nowak}:
if symbol $s_i$ is used to designate object $r_j$, then $A_{ji} =1$; otherwise
this element is zero.  The same symbol may be used to designate more than one object
(in principle, all $m$ objects could be designated by the same symbol), and
several symbols may refer to the same object.  By definition, each object is represented
by at least one symbol, so that
each row of {\sf A} possesses at least one nonzero element.

A key assumption of the model is that in the communication between speaker and
listener, all objects occur with the same probability, so that $p(r_j) = 1/m$ for
all $j$.  (In a subsequent work \cite{FC} this assumption was relaxed; the case of equally likely
objects is called ``model B" in \cite{FCDZ}.  Since Prokopenko et al. \cite{PAOP} argue that
the behavior of model A, with unequal object probabilities, is essentially the same, we focus here
on the simpler model B.)
Define a {\it communication event} as the occurrence of an object and
the speaker reporting this to the listener.
When the speaker refers to object $r_j$, she uses each of the symbols
that refer to this object (i.e., those for which $A_{ji} = 1$) with equal likelihood.
This implies that the probability of symbol $s_i$ over the space of all possible
communication events is

\begin{equation}
p(s_i) = \sum_{j=1}^m p(s_i,r_j) = \sum_{j=1}^m p(r_j) p(s_i|r_j)
= \frac{1}{m} \sum_{j=1}^m \frac{A_{ji}}{\sum_{k=1}^n A_{jk}}
\equiv \frac{1}{m} \sum_{j=1}^m B_{ji}
\label{psi}
\end{equation}

\noindent where we have introduced matrix {\sf B}, obtained from {\sf A}
by dividing the elements of each row by the corresponding row sum.
(Thus the row sums of {\sf B} are all unity.  Note that $B_{ji}$ is equal to the
conditional probability $p(s_i|r_j)$.)

The speaker chooses among $n$ symbols, from a probability distribution $p(s_i)$.
Following Shannon \cite{shannon}, the
mean information per symbol may therefore be defined as

\begin{equation}
H_n({\cal S}) \equiv - \sum_{i=1}^n p(s_i) \log_n p(s_i)
=-\frac{1}{\ln n} \sum_{i=1}^n p(s_i) \ln p(s_i) \equiv \alpha
\label{alpha}
\end{equation}

\noindent Note that the use of $\log_n$ imposes the condition $0 \leq \alpha \leq 1$.
Ferrer i Cancho and Sol\'e
interpret this quantity as the {\it speaker's cost} in communicating.  It is zero when only one
symbol is used (an impoverished language indeed!) and unity when all $n$ symbols have the same
probability.

The {\it listener's cost}, $\beta$, is related to uncertainty; when each symbol refers to a unique
object, there is no uncertainty and $\beta = 0$.  To define the listener's cost in the presence of
uncertainty, we begin by defining the cost in interpreting symbol $s_i$:

\begin{equation}
H_m ({\cal R}|s_i) \equiv - \sum_{j=1}^m p(r_j|s_i) \log_m p(r_j|s_i)  \equiv h_i,
\label{hrsi}
\end{equation}

\noindent where the conditional probability of object $r_j$, given reception of symbol $s_i$, is

\begin{equation}
p(r_j|s_i) = \frac{p(r_j)}{p(s_i)} p(s_i|r_j) = \frac{B_{ji}}{m p(s_i)}
= \frac{B_{ji}}{\sum_{k=1}^m B_{ki}} \equiv C_{ji} .
\label{defC}
\end{equation}

\noindent In the final equality we have defined {\sf C} as the matrix obtained from {\sf B} by dividing
each element in column $k$ by the corresponding column sum, so that each column sum
in {\sf C} is unity.  Thus,

\begin{equation}
h_i = - \frac{1}{\ln m} \sum_{j=1}^m C_{ji} \ln C_{ji}.
\end{equation}

\noindent The cost per symbol to the listener is then defined as

\begin{equation}
H_m ({\cal R}|{\cal S}) = \sum_{i=1}^n p(s_i) h_i \equiv \beta.
\label{beta}
\end{equation}

\noindent Evidently, $\beta$ is also restricted to [0,1].  Both $\alpha$ and $\beta$ are invariant
under permutations of the $n$ symbols, and of the $m$ objects.

Ferrer i Cancho and Sol\'e define the {\it total cost} as the linear combination:

\begin{equation}
\Omega(\lambda) = \lambda \beta + (1-\lambda) \alpha.
\label{Omega}
\end{equation}

\noindent Small values of the parameter $\lambda$ place a larger emphasis on the speaker's cost and vice-versa.
The authors of \cite{FCS} study the problem of minimizing $\Omega (\lambda)$ numerically,
and report that at a certain
critical value, $\lambda_c \simeq 0.41$, a phase transition occurs, at which certain properties such as the
effective lexicon size,
change in a singular manner.  (In Ref.~\cite{FCDZ} this conclusion was revised to reflect that the
transition actually occurs at $\lambda=1/2$.)

In references \cite{FCDZ,trosso,PAOP} the {\sf A} matrices minimizing $\alpha$ are identified as those in which all
nonzero elements fall in the same column. i.e., the speaker uses only a single word, so that $\alpha=0$.
These references also show that
to minimize $\beta$ in the symmetric case $m=n$, each column of {\sf A} must have only one nonzero element;
in this case there is no uncertainty and $\beta=0$.  Matrices with this property correspond to the
unit matrix and row permutations thereof.
These two classes of matrices minimize $\Omega (\lambda)$ for $\lambda < 1/2$ and $\lambda > 1/2$, respectively.
For $\lambda = 1/2$,
the class of minimizing matrices is larger, encompassing all those with $\alpha + \beta = 1$,
which implies that each row of {\sf A} has one and only one nonzero element, as shown below.
A consequence of these results is that

\begin{equation}
\min \Omega(\lambda) = \min \{\lambda, 1-\lambda\}.
\label{omega2fam}
\end{equation}

\noindent In the following section we demonstrate this result via direct calculation.

\section{Minimum cost states}\label{sec.states}

In this section we demonstrate Eq. (\ref{omega2fam}) for the case $n=m$ using a simple, direct approach.
We begin by showing that if $\alpha + \beta \geq 1$
for {\it any} matrix {\sf A}, then Eq. (\ref{omega2fam}) follows.  To see this, note that
$\alpha + \beta \geq 1$ implies that

\begin{eqnarray}
\lambda \beta + (1-\lambda) \alpha &\geq& \lambda \beta + (1-\lambda) (1-\beta)
\nonumber
\\
&=& (2 \lambda -1)\beta + 1 - \lambda
\nonumber
\\
&\geq& 1-\lambda, \;\;\; \mbox{for } \lambda \geq 1/2,
\end{eqnarray}
\noindent and
\begin{eqnarray}
\lambda \beta + (1-\lambda) \alpha &\geq& \lambda (1-\alpha) + (1-\lambda) \alpha
\nonumber
\\
&=& (1- 2 \lambda)\alpha + \lambda
\nonumber
\\
&\geq& \lambda, \;\;\; \mbox{for } \lambda \leq 1/2.
\end{eqnarray}

\noindent
Since there {\it are} matrices {\sf A} which render $\alpha = 0$ and $\beta=1$,
and others for which $\alpha=1$ and
$\beta = 0$, we know that it is in fact possible to saturate the inequalities, i.e.,
to have $\Omega = \lambda$ for $\lambda < 1/2$, and $\Omega = 1-\lambda$ for $\lambda > 1/2$.
Thus, if we can prove $\alpha + \beta \geq 1$, we will have established Eq. (\ref{omega2fam}).

It is not difficult to demonstrate the inequality.
Let {\sf B} be an $n \times n$ matrix with the following properties:

(i) $B_{ji} \geq 0$;\\
\hspace*{1em} (ii) each row contains a nonzero element;\\
\hspace*{1em} (iii) each row sum is unity: $\sum_{\ell = 1}^n B_{k \ell} = 1$.

\noindent We note that property (ii) corresponds to the rule that
each object must be represented by at least one symbol, and
that (iii) reflects normalization of the $B_{ji}$, which, for each $j$, constitute a conditional
probability distribution.
Define {\sf C} as the matrix obtained from {\sf B} by dividing the elements in each (nonzero) column
by the corresponding column sum.  Evidently {\sf B} and {\sf C} correspond to the matrices obtained from
the lexical matrix {\sf A}, as defined in Sec. II.  In a simplified notation let $p_i \equiv p(s_i)$ denote
the probability of symbol $s_i$, so that for each $i = 1,...,n$, we have:

\begin{equation}
p_i \equiv \frac{1}{n} \sum_{k=1}^n B_{ki}.
\end{equation}

\noindent Then $p_i \geq 0$, and $\sum_{i=1}^n p_i = 1$ by property (iii).  Define

\begin{eqnarray}
\alpha &\equiv& -\frac{1}{\ln n} \sum_{i=1}^n p_i \ln p_i
\nonumber
\\
       &=& -\frac{1}{n \ln n} \sum_{i=1}^n \left( \sum_{k=1}^n B_{ki} \right)
       \left[\ln \sum_{\ell = 1}^n B_{\ell i} - \ln n\right]
\nonumber
\\
       &=& 1 - \frac{1}{n \ln n} \sum_{i=1}^n
       \left(\sum_{k=1}^n B_{ki} \right) \ln \sum_{\ell = 1}^n B_{\ell i}.
\label{alpha1}
\end{eqnarray}

Next, let

\begin{equation}
h_i \equiv -\frac{1}{\ln n} \sum_{j=1}^n C_{ji} \ln C_{ji}
    = -\frac{1}{\ln n} \sum_{j=1}^n \left( \frac{B_{ji}}{\sum_{k=1}^n B_{ki}} \right)
    \left[\ln B_{ji} - \ln \sum_{\ell = 1}^n B_{\ell i} \right],
\end{equation}

\noindent and define $\beta = \sum_{i=1}^n p_i h_i$.  Using the expressions above for $p_i$ and $h_i$, we find

\begin{eqnarray}
\beta &=& -\frac{1}{n \ln n} \sum_{i=1}^n \left( \sum_{r=1}^n B_{ri} \right)
          \sum_{j=1}^n \left( \frac{B_{ji}}{\sum_{k=1}^n B_{ki}} \right)
          \left[\ln B_{ji} - \ln \sum_{\ell = 1}^n B_{\ell i} \right]
\nonumber
\\
      &=& -\frac{1}{n \ln n} \sum_{j=1}^n \sum_{i=1}^n B_{ji}
      \left[\ln B_{ji} - \ln \sum_{\ell = 1}^n B_{\ell i} \right],
\end{eqnarray}

\noindent so that

\begin{equation}
\alpha + \beta = 1 - \frac{1}{n \ln n} \sum_{j=1}^n \sum_{i=1}^n B_{ji} \ln B_{ji}.
\label{aplusb}
\end{equation}

\noindent For each $j$, $\sum_{i=1}^n B_{ji} = 1$, and since $n$ and the $B_{ji}$ are nonnegative,
$-\sum_{i=1}^n B_{ji} \ln B_{ji} \geq 0$. $\blacksquare$

Thus Eq. (\ref{omega2fam}) represents the global minimum of $\Omega(\lambda)$.
At $\lambda=1/2$,
there is a ``phase transition", or better, a change in the nature of the ground state,
at which the number $L$ of words used jumps from 1 to $n$.  (In this sense, the transition is
{\it discontinuous}.)

\subsection{Multiplicity and nature of minimum-cost states}

From the results cited at the end of Sec. II, it is evident that (for $n=m$)
there are $n$ matrices {\sf A} for which $\alpha = 0$, and
$n!$ matrices such that $\beta = 0$.
As noted, for a matrix {\sf A} to satisfy $\alpha + \beta = 1$, it must have one and only one nonzero
element in each row.  (This follows from Eq.~(\ref{aplusb}): if any row sum were greater than unity,
the nonzero elements in the corresponding row of {\sf B} would be smaller than unity, making $\alpha + \beta$
strictly greater than 1.)  The number of matrices that
minimize $\Omega(\lambda =1/2)$ is therefore $n^n$.  Thus the multiplicity
of the minimum-cost state is different for $\lambda < 1/2$, $\lambda > 1/2$, and $\lambda=1/2$.
Denoting the multiplicity by ${\cal M}(\lambda)$, we have
${\cal M}(\lambda > 1/2)/{\cal M}(1/2) \simeq e^{-n}$ and
${\cal M}(\lambda < 1/2)/{\cal M}(1/2) = 1/n^{n-1}$.  The extremely small value of the latter ratio
may be related to the observation \cite{FCDZ,PAOP} that a greedy search algorithm has difficulty finding
global minimum-cost states for $\lambda < 1/2$.

We turn now to the rank-frequency relation in minimum-cost states.
For a given matrix {\sf A}, rank the symbols in order of decreasing probability, and let $p_k$
be the normalized frequency of the $k$-th symbol in the ranking.
Now consider, for a given size $n$ (with $n=m$, as before), the set of matrices {\sf A} that
minimize $\Omega(\lambda)$; all matrices in this set are assigned the same probability.
Of principal interest is the mean,
$P(k) \equiv \langle p_k \rangle_\lambda$, over all matrices minimizing $\Omega(\lambda)$.
For $\lambda < 1/2$, $ P(k) = \delta_{k,1}$ while for $\lambda > 1/2$, there is no ``ranking"
as each symbol has the same probability, $1/n$.  Thus the rank-frequency relation is uninteresting
for $\lambda \neq 1/2$.

\begin{figure}[h]
\begin{center}
\includegraphics[height=6.5cm,width=7.4cm]{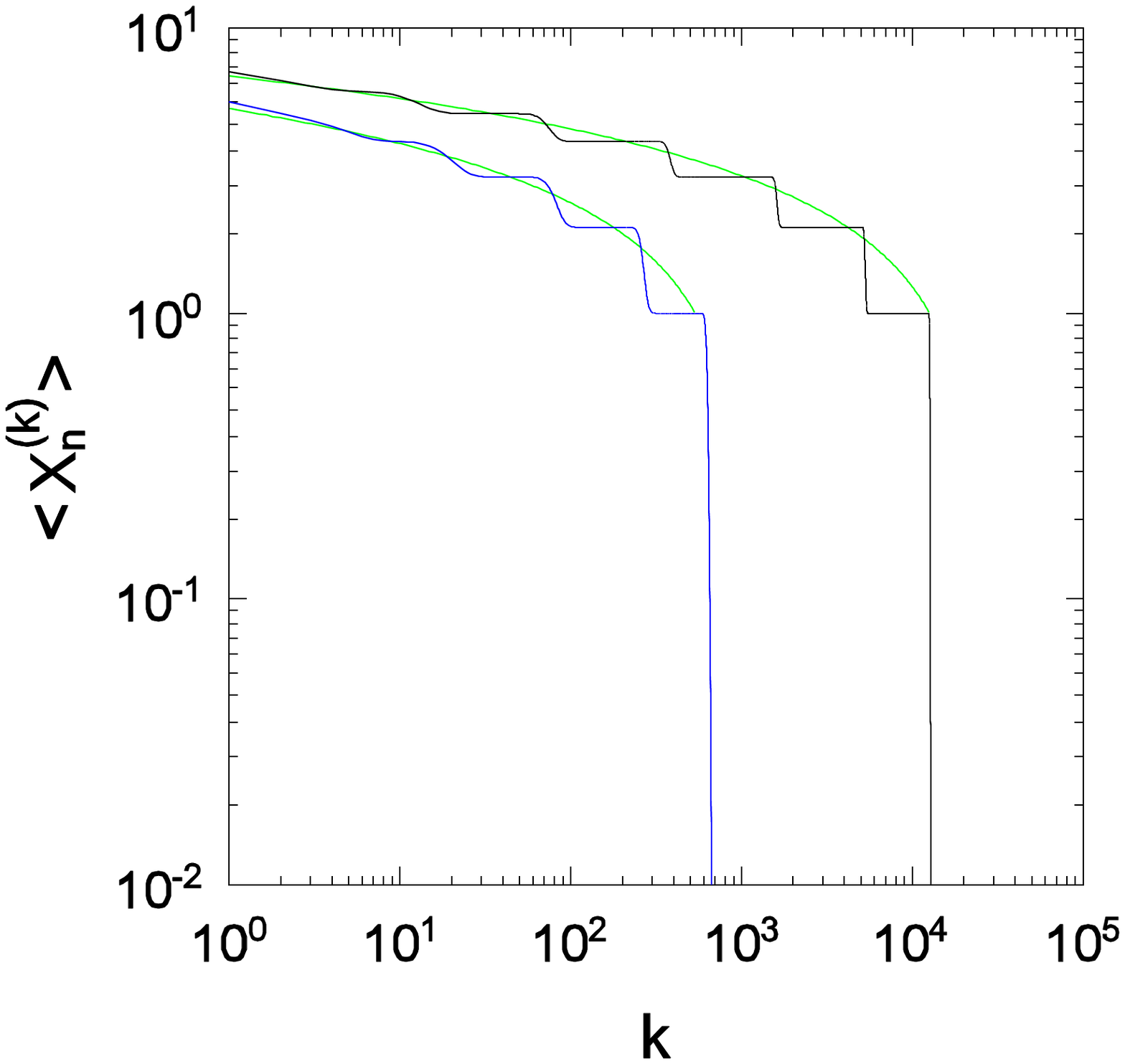}
\includegraphics[height=6.5cm,width=7.4cm]{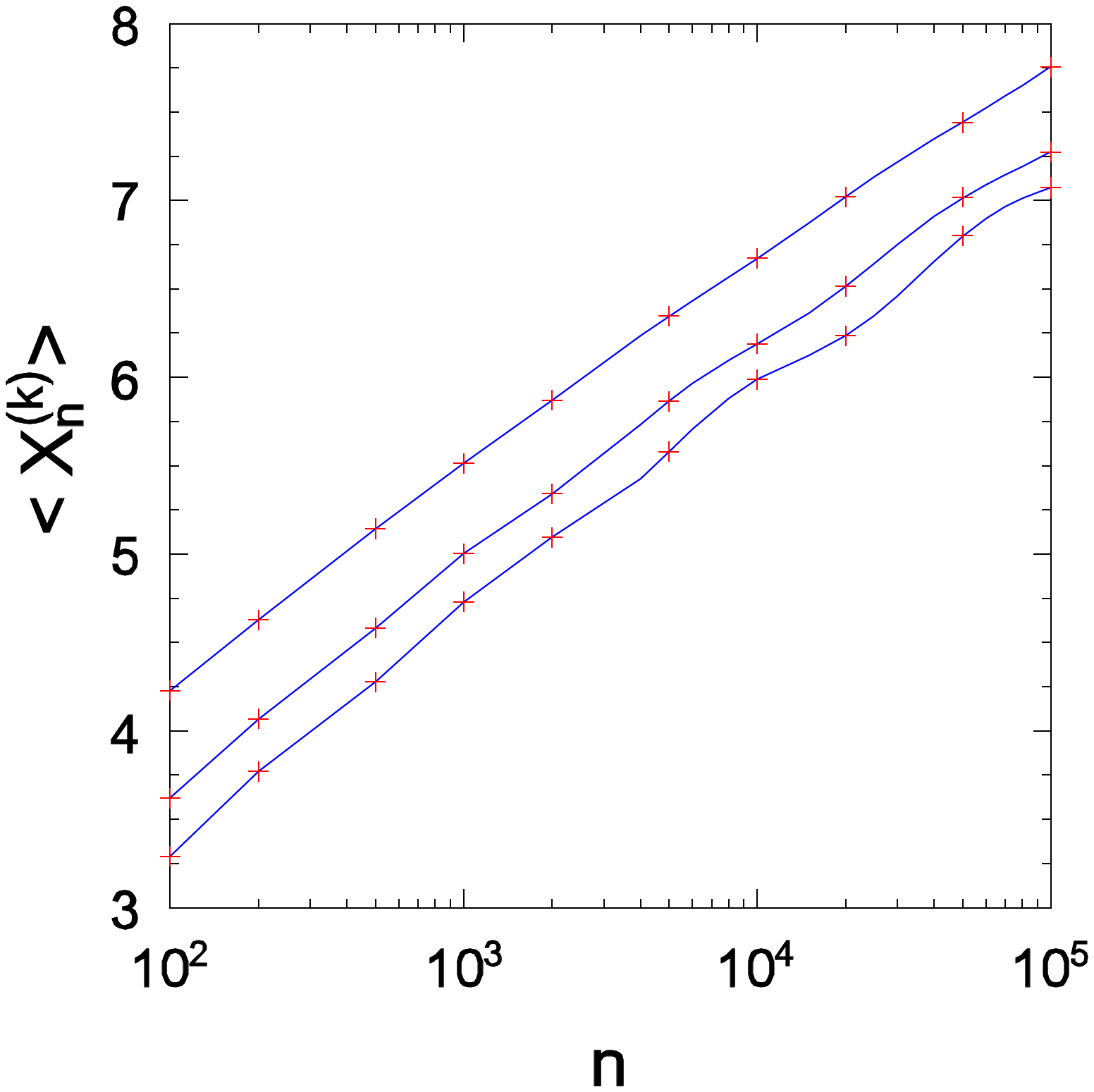}

\vspace{-1em}

\caption{\footnotesize{Left: mean symbol frequency $\langle X_n^{(k)} \rangle$ versus rank $k$ for
$\lambda = 1/2$, obtained via simulation, for
$n=1\,000$ (blue curve), and $n=20\,000$ (black curve).  Data are averages over $2 \times 10^5$ realizations.
The smooth curves (green) are inverse-factorial distributions.
Right: Mean symbol frequency versus system size $n$ for (upper to
lower) $k=1$, 2, and 3, for
$\lambda = 1/2$, obtained using the exact expressions (blue curves, see Appendix, Eqs.(25)-(27)) and
via simulation (red points).
}}
\label{poi3}
\end{center}
\end{figure}

For $\lambda = 1/2$, the minimum-cost matrices are those having exactly one nonzero
element in each row; the positions of these elements within each row are arbitrary and mutually independent.
Thus the number $q_i$ of nonzero elements in {\it column} $i$ is a binomial random variable (RV) with parameters
$n$ and $p=1/n$, i.e., Prob[$q_i=r$] = $\binom{n}{r} (1/n)^r (1-1/n)^{n-r}$.  The $q_i$ are subject to the
constraint $\sum_{i=1}^n q_i = n$.  In the limit of large $n$ with $\langle q_i \rangle = np = 1$ fixed,
$q_i$ approaches a Poisson RV with parameter 1, and the single constraint linking $n$ variables becomes
unimportant.
Thus for $n$ large, the numbers
of nonzero elements in columns 1, 2,...,$n$, are essentially independent, identically
distributed Poisson random variables with parameter 1.
(The validity of this approximation is verified below in a numerical example.)
$P(k)$ is readily estimated via
simulation, which
consists in generating a set of $n$ independent Poisson deviates with parameter 1 and
sorting them into decreasing order, $X_n^{(1)}$, $X_n^{(2)}$,...,$X_n^{(n)}$.
For $\lambda=1/2$, $P(k)$ is
the average of $X_n^{(k)}/n$ over many such independent realizations.
The resulting rank-frequency distribution, shown in Fig.~\ref{poi3} (left panel),
is not a power law.
Even ignoring the precipitous fall for $X_n^{(k)} < 1$, which corresponds to signals that have less than
one connection on average, we see that the bulk of the distribution
is characterized by a series of integer-valued steps.  Moreover, a smooth function interpolating the
steps would evidently decay more rapidly than a power law.

Prokopenko et al. \cite{PAOP} address the issue of the rank-frequency relation at the transition in a
complementary manner, by first identifying the most probable set of word occurrences, $\{\pi_i\}$,
i.e., the set that can be realized in the largest number of ways.
Each minimum-cost $n \times n$ matrix corresponds to a
{\it partition} $\{\pi_i \}$ of $n$, that is, a set of nonnegative integers with
$\sum_{i=0}^n i \pi_i = n$.  ($\pi_i$ is the number of columns having exactly $i$ nonzero elements.)
As noted
in \cite{PAOP}, a given partition corresponds to

\begin{equation}
W(\{\pi_i \}) = \frac{(n!)^2} {\prod_{i=0}^n \pi_i! (i!)^{\pi_i}}
\label{Wpi}
\end{equation}
\vspace{.1em}

\noindent distinct matrices obtained via row and column permutations.
Prokopenko et al. show that the rank-frequency distribution associated
with the partition $\{\pi_i \}$ that maximizes $W$ approaches
(for large $n$) an {\it inverse-factorial law}, which in our notation may be written so:

\begin{equation}
\langle X_n^{(k)} \rangle \simeq \Gamma^{-1} \left( \frac{n e^{\xi-1}}{k} \right),
\label{invfac}
\end{equation}

\noindent where $\Gamma^{-1}$ denotes the inverse of the gamma function, $\Gamma(x) = (x-1)!$, and
$\xi$ is a parameter.  As Prokopenko et al. point out, Eq. (\ref{invfac}) is incompatible with a
power-law.  Our simulation data for $n=20\,000$ yield a rank of $k \simeq 35$ for
$\langle X_n^{(k)} \rangle = 5$, implying that $e^{\xi-1} \simeq 1.26$.
(A similar analysis for $n=1\,000$ yields $e^{\xi-1} \simeq 1.07$.)
The resulting inverse-factorial
distributions indeed provide good fits to our simulation data (see Fig. 1, right panel).  This
suggests that the properties of a {\it typical} (or most probable) {\sf A} matrix
are similar to those obtained via an {\it average} over all minimum-cost {\sf A} matrixes.
Some years before, Trosso found that, averaging over the set of $n \times n$ minimum-cost matrices
for $\lambda = 1/2$, the ratios $P(k)/P(k+1)$ for $k> n/2$ converge to the value $n^2$ in the
infinite-$n$ limit, excluding power-law behavior for the second half of the distribution \cite{trosso}.

The values of $\langle X_n^{(k)} \rangle$ can be
calculated exactly for $k=1$, 2, and 3 (see Appendix); the result agrees with simulation,
as shown in Fig.~\ref{poi3} (right panel);  $\langle X_n^{(k)} \rangle$ is seen to increase very slowly
with $n$.

The mean lexicon size at $\lambda = 1/2$ is $n$
times the probability $P(X>0)$, where $X$ is a Poisson random variable (RV) with parameter 1.
Thus for large $n$ and $\lambda=1/2$, we have $\langle L \rangle/n = 1-e^{-1} \simeq 0.6321$.
Since the probability of having exactly $r$ nonzero elements in a given column, $i$, is
$P_r = 1/(er!)$, the symbol frequencies follow $p_i = r/n$, where $r$ is again a Poisson RV with
unit intensity, for $r=0, 1, 2,...,n$.
Thus the expected number of symbols having a normalized frequency of $r/n$ is $nP_r$, so that
for large $n$, the speaker's cost is
\begin{eqnarray}
\alpha &\simeq& -\frac{1}{\ln n} \sum_{j=0}^\infty \frac{j}{ej!} \ln \left(\frac{j}{n}\right)
\nonumber
\\
       &=& 1 - \frac{\phi}{\ln n},
\label{alpasymp}
\end{eqnarray}
where
\begin{equation}
\phi \equiv \frac{1}{e} \sum_{j=2}^\infty \frac{j \ln j}{j!}
\label{defphi}
\end{equation}
is found numerically to be 0.5734028...  At the transition then, the mean value of the speaker's cost
tends slowly to unity as the number of objects $n$ tends to infinity.

We have seen that the rank-frequency distribution does not follow a power law
when we average over the set of lexical matrices minimizing $\Omega$ at $\lambda=1/2$.
Next we examine the likelihood that a minimum-cost matrix (i.e., one for which $\alpha + \beta = 1$)
follows Zipf's law.
A small fraction of the minimum-cost matrices do in fact follow
a Zipf distribution (which we assume here to be a power law with exponent -1,
as originally proposed by Zipf);
we call these {\it Z-matrices}.  One such matrix can be constructed as follows.
In column 1, let the first $f_1$ elements be unity and the remainder zero; then in column 2, let
the elements in rows $f_1 + 1$ up to $f_1 + f_1/2$ be unity, and all others zero.  Proceed in this
manner until $f_1$ columns have been populated with $f_1$,
$f_2 = [f_1/2]$,...,$f_j = [f_1/j]$,...,$f_{f_1} = 1$ nonzero elements,
leaving the remainder of the columns with only zeros.  (Here
[...] denotes the largest integer of its argument.)  By construction, the symbol frequencies
follow a Zipf distribution.  The number $n$ of objects is approximately

\begin{equation}
n \simeq f_1 \sum_{j=1}^{f_1} \frac{1}{j}
= f_1 \left[\ln f_1 + \gamma + {\cal O}\left(\frac{1}{f_1} \right) \right],
\end{equation}

\noindent where $\gamma \simeq 0.5772$ denotes the Euler-Mascheroni constant.

The number $N_Z(n)$ of Z-matrices
is given by the number of choices of columns and rows.
Assuming that the $f_j$ are all distinct, we have $n!/(n-f_1)!$ choices for
the columns.  (Since there will, in general, be several columns with only one, or two, etc.,
nonzero elements, this is actually an overestimate.)  Independent of the column permutations,
we may permute the rows; the number of such permutations is approximately
$n!/[f_1! \,(f_1/2)! \,(f_1/3)! \cdots 2! \, 1!]$.  Using Stirling's formula one finds,

\begin{equation}
\ln N_Z(n) \simeq 2 n \ln n - (n-f_1) \ln (n-f_1) - f_1 - n \ln f_1 + f_1 \chi_{f_1},
\end{equation}
\noindent where
\begin{equation}
\chi_f \equiv \sum_{j=1}^f \frac{\ln j}{j}.
\end{equation}

Since the number $N_{MC}(n)$ of $n \times n$ matrices with $\alpha + \beta =1 $ is $n^n$,
the fraction represented by Z-matrices is extremely small.  For a Zipf distribution
of very modest length, $f_1 = 30$, one has $n=120$ and $N_Z/N_{MC} \simeq 10^{-41}$.
Increasing $f_1$ to 100 implies $n=519$, and the ratio becomes of order $10^{-313}$!
It is clear that
this tendency will not change even if we relax our requirements to consider a matrix
compatible with Zipf's law (e.g., by allowing the Zipf exponent to be $\mu \neq 1$, or
by allowing small fluctuations in the $f_k$ about a strict power law).
Thus, we conclude that
any reasonably sized Zipf distribution has essentially zero probability of
appearing in the set of minimum-cost matrices at $\lambda=1/2$.

\subsection{The case $\alpha=\beta$}

Matrices having $\alpha=\beta=1/2$ are of particular interest, as it has been
shown that (for $n\!=\!m$), power-law rank-frequency distributions (i.e., Z-matrices)
exhibit this property in the limit $n \to \infty$ \cite{PAOP}.  Although these authors noted that
there are significant finite-size corrections to this relation, the equal-cost criterion
seems worth investigating as a possible condition leading to Zipf's law.

We constructed Z-matrices as described in the preceding subsection, and evaluated the
costs $\alpha$ and $\beta$.  Studying matrices with $n$ in the range 100 - 10$^8$, we find
$\alpha$ values ranging from 0.58 to 0.55, with an apparent ($n \to \infty$) limit of about 0.51.
The fact that finite Z-matrices have $\alpha$ somewhat greater than $\beta$ is in qualitative
agreement with the results of \cite{PAOP}; that the apparent limit for $\alpha$ is $> 1/2$ may
be attributed to very slow convergence as $n \to \infty$.

Prokopenko et al. showed that (asymptotically) $\alpha = \beta$ is a {\it necessary} condition
for a power-law of the form $P(k) \propto 1/k$, but were unable to determine if this
represents a {\it sufficient} condition.  In fact it is not.  From Eq. (\ref{alpha1}) we have that,
for $\alpha = 1/2$,

\begin{equation}
\sum_{i=1}^n b_i \ln b_i = \frac{n \ln n}{2},
\label{Psi}
\end{equation}

\noindent where we defined the column sums of {\sf B} as $b_i \equiv \sum_{j=1}^n B_{ji}$.  (Note
that for the case considered here, $\alpha + \beta = 1$, matrices {\sf A} and {\sf B} are identical.)
If $n$ is a square number we can construct a non-Z-matrix with $\alpha = 1/2$
by placing $\sqrt{n}$ nonzero elements in
each of $\sqrt{n}$ different columns (with the remaining columns all zero), maintaining, as always,
exactly one nonzero element per row.  If $n$ is not square we can construct matrices with $\alpha \simeq 1/2$
by partitioning $n$ using a set of integers as near as possible to $\sqrt{n}$.  Thus $\alpha=\beta$
is not a sufficient condition for a power-law rank-frequency distribution.

\begin{figure}[h]
\begin{center}
\includegraphics[height=8cm,width=11cm]{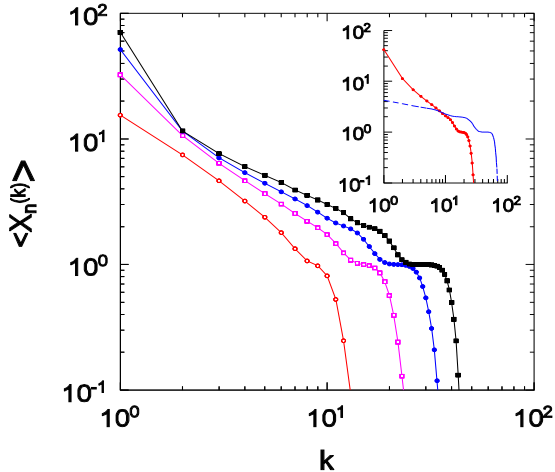}

\vspace{-1em}

\caption{\footnotesize{Mean symbol frequency $\langle X_n^{(k)} \rangle$ versus rank $k$ associated with
equal-cost matrices ($\alpha=\beta=1/2$) for system sizes (left to right) $n=40$, 80, 120, and 160
(colors red, violet, blue, and black, respectively).
Inset: comparison of the equal-cost distribution (red points) with
the average over {\it all} matrices with $\alpha+\beta = 1$, as in Fig. \ref{poi3} (broken blue line),
for $n=100$.
}}
\label{aeb1}
\end{center}
\end{figure}

Even if not all equal-cost matrices correspond to a power-law distribution, one may ask if,
on average, such matrices exhibit any interesting properties.  To investigate this issue
we study the rank-frequency distribution averaged over all minimum-cost matrices
with $|\alpha - \beta| < \epsilon$, for some reasonably small ``tolerance" $\epsilon$.
As noted above, each
partition $\{\pi_i \}$ of $n$, corresponds to a number $W(\{\pi_i \})$
of $n \times n$ minimum-cost matrices, given by Eq. (\ref{Wpi}).
We generate all partitions of $n$ and include those satisfying $|\alpha - \beta| < \epsilon$ in the average,
weighing each by the factor $W$.  The tolerance $\epsilon$ ranges from 0.02 for $n=40$ to $10^{-5}$ for $n=160$.
(The increasing selectivity is possible because of an extremely rapid growth in the number of
partitions with system size: for $n=40$ the sample includes 2250 partitions, while for $n=160$ there are
about $4.5 \times 10^6$, despite the much smaller tolerance.)
Rank-frequency distributions averaged over equal-cost
matrices are shown for several system sizes in Fig. \ref{aeb1}.
Again, a fast decay for $\langle X_n^{(k)} \rangle<1$ and a plateau at
$\langle X_n^{(k)} \rangle \approx 1$ are evident; these features are incompatible with Zipf's law.
However, neglecting this region, the distribution is more linear (on
log-scales) than that obtained from the unrestricted average (see
inset of Fig. 2).
It nevertheless appears unlikely that the rank-frequency distribution averaged over equal-cost
will yield a Zipf-like distribution: the decay exponent $\mu$ appears to decrease systematically
with system size; we find $\mu = 0.94(1)$ and $0.83(1)$ for $n=120$ and 160, respectively.
(The figure in parentheses denotes the uncertainty in the final
digit.  The $\mu$ values are obtained via least-square linear fits to $\ln \langle X_n^{(k)} \rangle$ versus
$\ln k$ over the largest interval that appears to be compatible with a power law, determined visually.)

\section{Simulation}\label{sec.simulation}

\begin{figure}
\includegraphics*[width=\columnwidth]{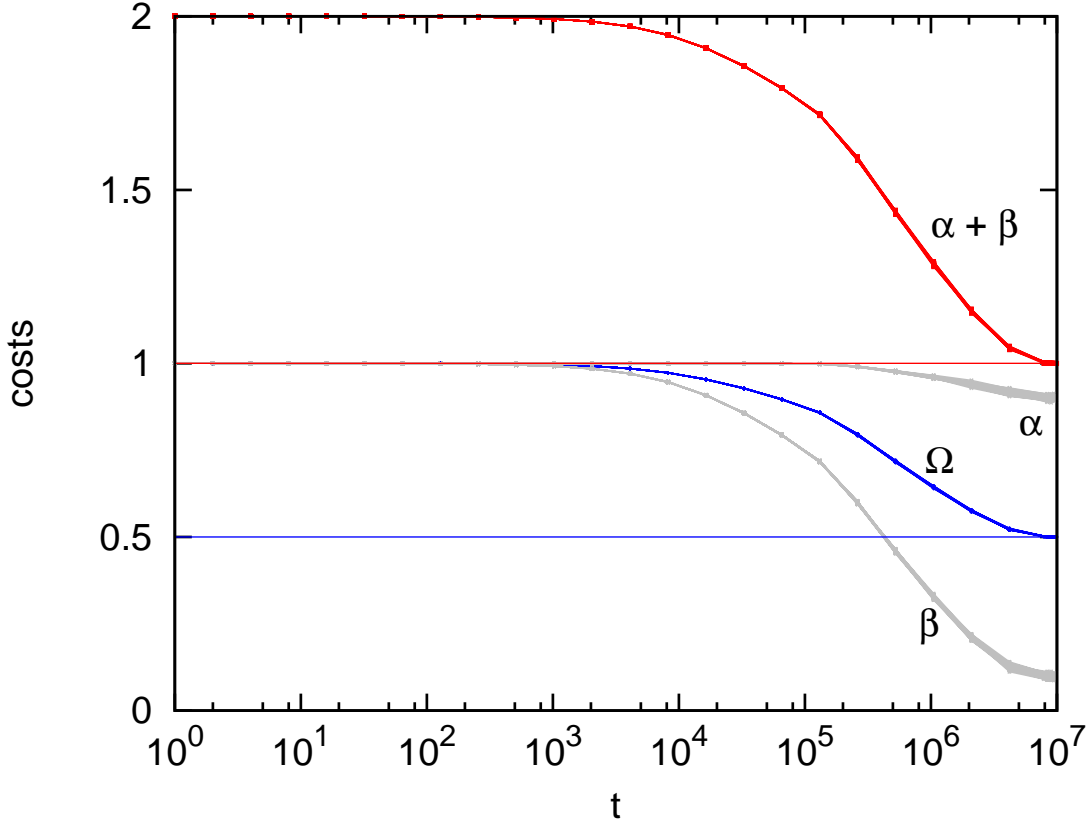}
\caption{Evolution of costs $\alpha$ and $\beta$ (grey curves),
$\Omega$ (blue) and $\alpha+\beta$ (red),
for $122$ independent simulations, for
  $\lambda = 1/2$ and $n=300$. Time is measured in Monte Carlo steps
  per matrix element. The horizontal lines at heights $1/2$ and $1$
  indicate the values theoretically minimizing $\Omega(\lambda)$ and
  $\alpha + \beta$, respectively.}
\label{F:cost_evol}
\end{figure}

Our aim in this section is to determine whether a minimum-search algorithm
is able to attain the minimum-cost states identified above.
(Note that the simulations reported in the preceding section do not involve
{\it searching} for the minimum cost; they are merely used to estimate the rank-frequency
relation for a set of independent Poisson RVs.)
We apply a Monte Carlo algorithm
as described in \cite{FCS} to the minimization of
$\Omega(\lambda)$: with probability $\nu$ an element of the lexical matrix
{\sf A} is flipped. If the resulting cost is lower, the flip is
accepted, otherwise it is rejected. (Naturally, flips from 1 to 0
that would leave a row with all elements zero are also rejected.)
This procedure is repeated a large
number of times (of the order of $10^7$ Monte Carlo steps per matrix
element). A flipping probability $\nu = 4/(n(n-1))$ is used as in
\cite{FCS}. To track the evolution of $\Omega(\lambda)$ over time we
initialize the matrix {\sf A} by setting all elements equal to
$1$. Typical evolutions are shown in Fig.~\ref{F:cost_evol} for
$\lambda=1/2$ and $n=300$, in which each of the 122
simulations in the ensemble have attained global minimum cost states
associated with $\Omega(1/2) = 1/2$.

\begin{figure}
\includegraphics*[width=\columnwidth]{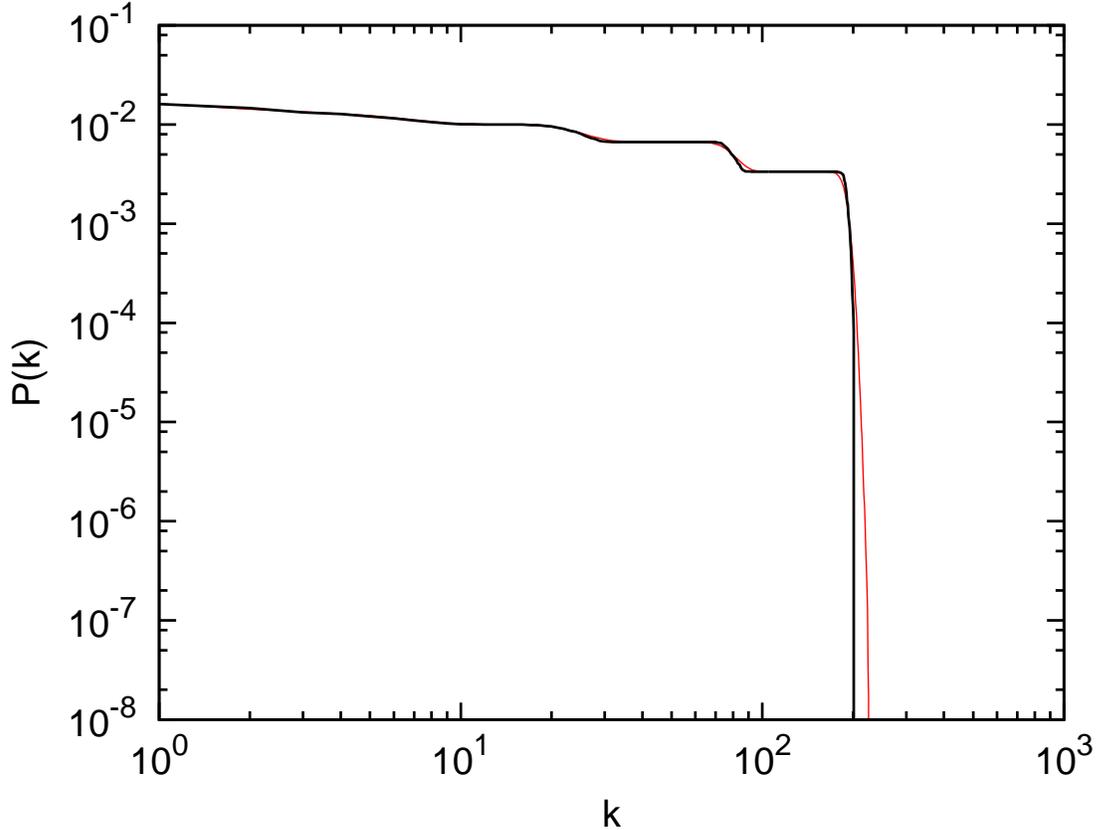}
\caption{(Black line) Normalized symbol frequency $P(k)$ versus rank $k$ for $\lambda = 0.5$
  and $n = 300$, averaged over $122$ simulations, using a spin-flipping
  algorithm as described in~\cite{FCS}. (Red line) $P(k)$ obtained
  from Poissonian statistics as described in Sec.~\ref{sec.states}.}
\label{F:pk}
\end{figure}

Clearly, little variation is seen. Indeed, the speaker's
cost is found to be $\langle \alpha \rangle = 0.901(5)$,
as compared to $\alpha = 0.8995$ obtained from
Eq.~\eqref{alpasymp} for $n=300$.  The histogram for the normalized symbol
frequency in Fig.~\ref{F:pk} for $n=300$ (black line) is very similar
to that obtained via the Poissonian statistics approach of the
preceding section (red line), as well as to Fig.~3C in \cite{FCS}, for
$n=150$.  This validates our analysis based on
independent Poisson RVs even for rather modest system sizes.

We note that the performance of the algorithm deteriorates for $\lambda
< 1/2$. Thus, while it is possible to fully minimize the cost at
$\lambda = 1/2$ within a reasonable amount of time, the same is not
true for $\lambda < 1/2$, in which case the resulting distributions of
$P(k)$ are sensitive to details such as the initial condition for
${\sf A}$ and the amount of time the simulation is allowed to run.

\newpage

\section{Conclusions}\label{sec.conclusions}

In summary, we study the minimization problem associated with the Ferrer i Cancho-Sol\'e
model via direct calculation and simulation. Our analysis furnishes an alternative derivation of
the minimum-cost formula, $\min \Omega(\lambda) = \min \{\lambda, 1-\lambda\}$, via
the inequality $\alpha + \beta \geq 1$.
The result for the minimum cost was shown in several previous works,
using different approaches \cite{FCDZ,trosso,PAOP}.
While this expression and other aspects of the minimum-cost state are singular at $\lambda = 1/2$,
we find no evidence for the power-law frequency-rank distribution reported in \cite{FCS}.
Minimizing the cost only yields Zipf's law for a small fraction of
minima at $\lambda=1/2$, which vanishes for the relevant case of
increasing system size.
Our results suggest that at the transition, the properties of a typical
cost-minimizing {\sf A} matrix, as determined in
\cite{PAOP}, are similar to the properties of an average {\sf A} matrix, as one
would expect for a statistical model free of quenched disorder.  In particular, the
inverse-factorial law derived in \cite{PAOP} for the typical minimum-cost matrix
also describes the envelope of the rank-frequency distribution averaged over all
such matrices.

It has been suggested that the Zipf-like distribution is associated with sub-optimal
states, with costs slightly greater than the minimum \cite{FCDZ}.  Our simulations, however, do not
yield a power-law distribution in this slightly sub-optimal situation, casting doubt on
whether Zipf's law can be explained in this manner.
These authors also speculated {\em ``that Zipf's law ... could be the
consequence of local minima of $\Omega(\lambda)$.''}
Our numerical simulations show that non-Zipfian
minimum-cost states are attained at the transition.

Finally, it is interesting to speculate about which alterations of the model
of Ref.~\cite{FCS} could lead to Zipf's law.
While models with substantial modifications have been investigated~\cite{FC,FCDZ,CMFS},
our finding of
states compatible with Zipf's law at $\lambda=1/2$ suggests
that small modifications of the model might be sufficient to break the degeneracy,
such that a power-law distribution would correspond to the global minimum cost.
One such possibility is the equal-cost criterion, shown in \cite{PAOP} to be (asymptotically)
a necessary condition for a Zipf distribution.  Although we have shown that this condition is not
sufficient, we also find that the average over all equal-cost matrices yields distributions that
are closer to a power law.  It appears unlikely, however, that this alone
is sufficient to generate Zipf's law.
It is also worth considering the possibility that the evolution of language cannot
reach the minimum-cost states on historic time scales, and instead
wanders in a space of sub-optimal configurations, for which Zipf's
principle does hold to good approximation.
\vspace{2em}

\noindent{\bf Acknowledgments}

We are grateful to Paolo Cermelli and Oliver Obst for helpful correspondence.
This work was supported by CNPq, Brazil.
\vspace{2em}

\noindent {\bf Appendix: Expectation of the maximum of $n$ independent random variables.}
\vspace{1em}

Consider an integer-valued random variable $Y$ with probability distribution
$p_m = \mbox{Prob}[Y\!=\!m]$, and let $q_m \equiv \mbox{Prob}[Y \! \leq \! m]$.
Let $Y_1$,...,$Y_n$, be a set of $n$ independent random variables drawn from
this distribution, and let $X_n^{(k)}$ denote the $k$-th largest variable in this set.
The event $X_n^{(1)}=m$ corresponds to having one or more of the $Y_i$ equal to
$m$, and all others smaller.  Thus

\begin{equation}
\mbox{Prob} [X_n^{(1)}=m] = \sum_{j=1}^n \binom{n}{j} p_m^j q_{m-1}^{n-j} = q_m^n - q_{m-1}^n,
\end{equation}

\noindent and $\langle X_n^{(1)} \rangle$ is given by the (conditionally convergent) sum

\begin{equation}
\langle X_n^{(1)} \rangle = \sum_{m=0}^\infty m[q_m^n - q_{m-1}^n].
\label{expxn}
\end{equation}

For the Poisson distribution with parameter unity, $p_m = 1/(e m!)$, the above sum converges
quite rapidly, with the contribution due to terms with $m \geq 20$ being negligible
for the $n$ values considered here.  Figure 2 shows that the simulations of Sec. III
are in good agreement with our analysis.

The means of the second and third largest variables can be obtained using,
\begin{equation}
\mbox{Prob} [X_n^{(2)}=m] = \mbox{Prob} [X_n^{(1)}=m]
      + n \left\{(1\!-\!q_m)[q_m^{n-1} - q_{m-1}^{n-1}] - p_m q_{m-1}^{n-1} \right\}
\end{equation}
\noindent and
\begin{eqnarray}
\mbox{Prob} [X_n^{(3)}\!=\!m] &=& \mbox{Prob} [X_n^{(2)}\!=\!m]
\nonumber
\\
      &+&\! \mbox{\small $\frac{n(n-1)}{2}$}
      \left\{(1\!-\!q_m)^2[q_m^{n-2} - q_{m-1}^{n-2}] -p_m q_{m-1}^{n-2}[p_m+2(1\!-\!q_m)] \right\}
\end{eqnarray}

\noindent Analogous formulas can of course be derived for the fourth and subsequent variables,
though they become increasingly more complicated.  We have verified that
the simulations agree with the exact expressions for the means of $X_n^{(1)}$, $X_n^{(2)}$, and $X_n^{(3)}$.
For example, for $n=1000$ we have $\langle X_n^{(k)} \rangle = 5.51384$, $5.00381$, and $4.73083$ for
$k=1$, $2$, and $3$, respectively, while simulation yields $5.5136(3), 5.0040(4)$ and $4.7308(6)$,
with the figures in parentheses denoting statistical uncertainties.

\bibliographystyle{apsrev}

\end{document}